\documentstyle[11pt]{article}
\begin{document}

\pagestyle{empty}
\baselineskip 0.2in
\begin{titlepage}

\vspace{0.2in}
\rightline{\vbox{\halign{&#\hfil\cr
&NTUTH-97-02\cr
&January 1997\cr}}}

\vfill

\begin{center}
{\Large \bf  THE CASE FOR A VECTOR GLUEBALL
\footnote
{
Talk presented at the YITP International Workshop 
``Recent Developments in QCD and Hadron Physics",
Kyoto, Japan, December 16--18, 1996. 
To appear in Proceedings.
}
}
\vfill
        {\bf George Wei-Shu HOU}
\footnote{
E-mail: wshou@phys.ntu.edu.tw.} \\

        {Department of Physics, National Taiwan University,}\\
        {Taipei, Taiwan 10764, R.O.C.}\\
\end{center}
\vfill
\begin{abstract}
We argue that the $1^{--}$ glueball $O$ is much cleaner
than other glueballs candidates.
Anomalously large $J/\psi \to \rho\pi,\ K^*\bar K$ decays
and the scalar glueball mass scale
suggest that $m_{O} \simeq m_{J/\psi}$.
\end{abstract}
\vfill
\end{titlepage}

\begin{center}

{\large\bf  THE CASE FOR A VECTOR GLUEBALL }\\[.25in]

{\bf  George Wei-Shu Hou} \\

Department of Physics,
National Taiwan University\\
Taipei, Taiwan 10764, R.O.C.

\end{center}

%\vglue.2in
\begin{abstract}
We argue that the $1^{--}$ glueball $O$ is much cleaner
than other glueballs candidates.
Anomalously large $J/\psi \to \rho\pi,\ K^*\bar K$ decays
and the scalar glueball mass scale
suggest that $m_{O} \simeq m_{J/\psi}$.
\end{abstract}

\vskip 0.4cm
\noindent{\Large\bf 1. 
Introduction}
\vskip 0.1cm

As {\it bound states of gauge particles}, glueballs are unique and fundamental. 
Their existence would offer a direct proof of the nonabelian nature of QCD,
since there is no analog in QED,
while for $W$ and $Z$ bosons, analogous states are spoiled by
spontaneous symmetry breaking.

Unfortunately, it has been extremely difficult to establish such states, 
despite over 20 years of diligent search. 
Traditional search methods are: 
1) Glue-rich channels: e.g. in $J/\psi \to \gamma + gg \to \gamma + X$;
2) Classification: extra isosinglet mesons that do not fit into $q\bar q$ nonets;
3) Veto: absence in $\gamma\gamma$ production.
Fifteen years ago, the leading candidates were 
the $\iota(1440)$ and $\theta(1640)$.
But  the $\iota$ is now labeled the $\eta(1440)$
hence a $q\bar q$ meson,
while the old $\theta$ turned into the $f_J(1710)$,
and has recently been resolved by BES to consist of 
$0^{++}$ and $2^{++}$ components.

The $0^{++}$ glueball $G$, originally expected to be around 1 GeV, was 
problematic on the theory side since it has vacuum quantum numbers.
The experimental situation was also murky for a long time.
Recent progress \cite{Landua} in $p\bar p$ annihilation studies, however,
suggest that either the $f_0(1500)$ or $f_0(1710)$
could be the $0^{++}$ scalar glueball.
This is further supported by lattice results 
and a recent K-matrix fit to various data.
Although the situation seems robust,
the main difficulty continues to be how to disentangle the actual glue state 
from $q\bar q$ states.
Another long standing candidate is the $\xi(2230)$,
which is now observed in many channels by BES via method 1).
Unfortunately, statistics does not allow a spin/parity analysis,
although $2^{++}$ is favored.

How can the ``classification" method, even with lattice support, 
ever be conclusive?
We note the main culprit is the existence of {\it light} $q\bar q$ states:
isosinglet $q\bar q$ states in 1 -- 2 GeV region could easily mix with glueballs,
the fruit of pure Yang-Mills QCD.
This actually springs from an ``accident" in Nature --- the existence of
global SU$_{\rm F}$(3).
To imitate Nature, lattice calculations have to go un-quenched,
which means not only incorporating $q\bar q$--$G$ mixings,
but also the modification of scale due to quark loops.

%Can ``clean" glueballs be found?

\vskip 0.6cm
\noindent{\Large\bf 2. 
Can ``Clean" Glueballs be Found?}
\vskip 0.1cm

Since disentangling the glue from the quark states 
has been the main difficulty in the past,
by ``cleanness" we mean having small admixtures of $q\bar q$.
There are two (quantum mechanical) aspects:
\begin{description}
\item[\ \ \ \ Proximity:]  the density and distribution of states;
\item[\ \ \ \ Mixing Strength:] small coupling.
\end{description}
Small mixing is possible {\it IF} $m_{q\bar q}$ and $m($glueball) are 
{\it very different}, and, hopefully the latter is already in 
the asymptotic freedom regime.
Paradoxically, then, perhaps some low-lying but {\it heavy} glue state could be ``clean", 
as we try to demonstrate below.

An important question is the glueball mass scale.
The successful na\"\i ve quark model asserts
that light quarks have constituent mass of order 300 MeV,
which can be argued hand-wavingly from quark confinement
and dynamical chiral symmetry breaking.
Although there is no good analog of chiral symmetry 
(the gauge symmetry is local),
we wish to hand-wavingly argue for a ``constituent" gluon model
by analogy: gluons are quantized in a hadron ``box",
the size of which is determined dynamically,
but could be different from $q\bar q$ hadrons.
One can thus view gluons inside a glueball as having an
{\it effective mass $m_g$}, and move nonrelativistically, 
much like in the quark model.
As such, one can {\it count} the number of constituent gluons, 
as well as construct a potential model from 
$g$--$g$ scattering and gluon ``confinement".
Such a model was discussed by Cornwall and Soni \cite{CS, HS} a long time ago.
The glueball spectrum emerging from this model
is quite similar to various other approaches.
The lightest two gluon states are $0^{++}$ and then $0^{-+}$ and $2^{++}$.
For 3-g states, the four lowest lying ones are
$0^{-+}(1)$, $1^{--}(0)$,  $1^{--}(2)$, and a $3^{--}(2)$,
where numbers in parenthesis indicate total spin of any gluon pair,
and thus exchange symmetries.
We shall label these states as $G$, $P_2$, $\xi$, 
$P_3$, $O$ (two states) and $T$.
These are all lowest $S$-states with no orbital or radial excitation.
Note that the $1^{--}$ glueball cannot be made of two gluons.
The ratios $m_{\xi}/m_G$ and $m_O/m_G$ are found to
be $\sim 1.4$, $2.1$ respectively,
and all the 3-g states are found to be rather degenerate \cite{HS}.
Identifying $\theta(1640)$ as the $2^{++}$,
it was found that $m_G \sim 1.15$ GeV and $m_O \simeq 2.4$ GeV.
It was already noted that the heaviness of $O$ may
have interesting consequences for charmonium physics.

If we now take the currently favored value of $m_G\simeq 1.5$ GeV,
which is considerably higher than what people had thought,
we find $m_{\xi} \simeq 2.1$ GeV, close to BES observation,
and $m_O \simeq 3.1$ GeV, which is {\it very} close to $J/\psi$ mass
and {\it very heavy} for a pure glue hadron 
(the scale of QCD generally taken as $\sim$ 1 GeV)!
Although the situation in  $0^{++}$ and $2^{++}$ sector 
looks rather promising, as has been remarked, 
it may take a long time to disentangle them from quark states. 
On the other hand, the 3-g glueballs,
especially the $O$ state(s), seem to fit our bill for ``cleanness",
i.e. small mixing with $q\bar q$ states.
This can be argued from both its structure --- 3 constituent gluons ---
and its heaviness. The structure makes it difficult to convert 
to a $q\bar q$ configuration. Regarding its heaviness,
not only we have a paucity of nearby quark states, 
but also we {\it are in the asymptotic freedom domain}.

\vskip 0.6cm
\noindent{\Large\bf 3. 
The Case for a Vector Glueball}
\vskip 0.1cm

Let us investigate the expected $q\bar q$ content of the vector glueball $O$.
The lowest lying $1^{--}$ isoscalar $q\bar q$ mesons $V^0$
are $\omega$, $\phi$ and $J/\psi$, which are known to be
predominantly $n\bar n$ ($n = u,\ d$), $s\bar s$ and $c\bar c$ in $1S$ state.
With $m^2_\omega,\ m^2_\phi \cong 0.61,\ 1.02$ GeV$^2$ respectively,
they are very far from the lowest lying  $1^{--}$ glueball state $O$
with $m^2_O \sim 9$ GeV$^2$.
Furthermore, compared to the relative ease of
$G\to gg\to q\bar q$ mixing (helped by a lower scale!),
$V^0$-$O$ mixing occurs via the diagram,

%%Begin InstantTeX Picture
\let\picnaturalsize=N
\def\picsize{4.0in}
\def\picfilename{O_YITP.eps}
%If you do not have the picture file add:
%\let\nopictures=Y
%to the beginning of the file.
\ifx\nopictures Y\else{\ifx\epsfloaded Y\else\input epsf \fi
\global\let\epsfloaded=Y
\centerline{\ifx\picnaturalsize N\epsfxsize \picsize\fi \epsfbox{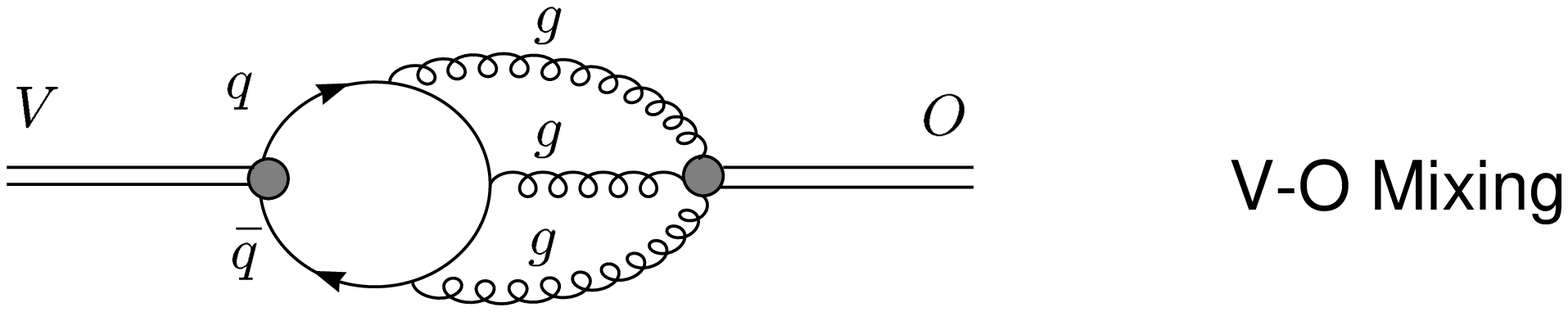}}}\fi
%%End InstantTeX Picture
%
\noindent hence there are three powers of the strong coupling constant,
{\it at a very high scale}.
Thus, with the help of asymptotic freedom,
this forms the basis for the smallness of OZI-violation
in vector mesons,
as exemplified in the near-ideal $\omega$-$\phi$ mixing.
In comparison, for scalar, pseudoscalar and tensor mesons,
the glueball states $G$, $P_2$ and $\xi$ are sufficiently close to
$f_0(1370)$-$f_0(1500)$-$f_0(1710)$,
$\eta$-$\eta^\prime$ and $f_2(1270)$-$f_2^\prime(1525)$,
such that singlet-octet mixing is far less than ideal,
with the slight exception for tensor mesons,
which is again due to the heaviness of $\xi$.

Upon taking $c\bar c$ mesons into account,
because of $m_O\simeq J/\psi$,
we expect that $J/\psi$ decays are likely affected,
which turns out to be supported by provocative experimental results.
We wish to note here that asymptotic freedom still leads to small
$O$-$J/\psi$ mixing, despite their proximity.
The state $P_3$ is also expected to affect the properties of $\eta_c$
because of proximity in mass. However, because of
$P_2$-$P_3$ mixing (2g--3g mixing),
it is expected to be less clean.

For sake of space, we can only briefly \cite{Hou} summarize the case for $O$:
\vskip 0.2cm
\noindent{\bf \boldmath{$\rhd$} History of $O$:} 

The state $O$ was proposed by Freund and Nambu (FN) in 1975,
to mediate OZI violating $\phi\to \rho\pi$ decay via the pole sequence
$\phi \to O \to \omega$. $O$ mass was argued to be $\sqrt{2} - \sqrt{3}$
GeV from dual dynamics, and prediction was made for 
$\Gamma(J/\psi\to \rho\pi)$, which turned out to be too large.

In 1982, Hou and Soni (HS) found \cite{HS} $m_O \simeq 2.4$ GeV from
potential (constituent gluon) model, which is much larger than FN.
To account for observed $J/\psi \to \rho\pi$ width,
HS introduced scale-dependent $O$-$V$ mixing
$f_{O\omega} : f_{O\phi} : f_{O\psi} = (\sqrt{2} : -1 : 1)f(q^2)$,
hence $f(m_{J/\psi}^2) \ll f(m_{\omega,\phi}^2)$ which suppresses $J/\psi\to \rho\pi$.
Note that BR$(J/\psi\to \rho\pi) \simeq 1\%$ is still quite prominent.
What was (and still is!) intriguing was the complete absence of
$\psi^\prime \to \rho\pi$, $K^*\bar K$ decay modes, 
as originally observed by MARK II.
Normally one expects
$R_{\psi^\prime/\psi}(X) \equiv B(\psi^\prime\to X)/B(J/\psi\to X)
                     \simeq B(\psi^\prime\to e^+ e^-)/B(J/\psi\to e^+ e^-)
                     \cong  0.15$,
often referred to as the ``15\%" rule, which is obeyed by most
$J/\psi$ and $\psi^\prime$ decay modes.
From the basis that $O$ is rather heavy, 
HS proposed a resonance enhancement model that
enhanced $\Gamma(J/\psi\to \rho\pi)$, while
$\Gamma(\psi^\prime\to \rho\pi)$ is untouched because of large
energy denominator.
Using this argument, the data at that time suggested
$m_O > 2.3$ GeV, which was quite consistent with potential model results.

By 1986, MARK III data had strengthened earlier findings of suppression
of $R_{\psi^\prime/\psi}(X)$ for $X = \rho\pi$, $K^*\bar K$.
To continue the ``Ansatz" of HS, 
Brodsky, Lepage and Tuan (BLT) proposed \cite{BLT} that
$O$ and $J/\psi$ were in fact close to being degenerate, i.e. 
\begin{equation}
\vert m_O - m_{J/\psi} \vert < 80\ \mbox{MeV},\ \ \ \
\Gamma_O < 160\ \mbox{MeV}.
\end{equation}
Besides following the HS ``Ansatz",
BLT also offered an argument for {\it suppression} of
vector $Q\bar Q$ states decaying into vector-pseudoscalar final state,
because of  ``hadron helicity conservation".

\vskip 0.2cm
\noindent{\bf \boldmath{$\rhd$} Recent Progress:}

In 1996, conflicting signals on the issue were given by the BES Collaboration.
On one hand, the $R_{\psi^\prime/\psi}(\rho\pi)$ anomaly seems to 
have deepened:
Not only the $VP$ modes $\psi^\prime \to \rho\pi$, $K^*\bar K$ remain
unseen down to the $10^{-5}$ level, it now appears that
$VT$ modes such as $\omega f_2$ etc. are also similarly anomalous.
On the otherhand, BES performed an energy scan \cite{scan} of the
$J/\psi\to \rho\pi$ mode and did not see any nearby state $O$,
hence the HS/BLT Ansatz seemed to be directly challenged.

On closer inspection \cite{Hou},
it turns out the scan result of ref. \cite{scan} was misinterpreted.
In short, 
$e^+e^-\to \rho\pi$ can be viewed as proceeding via
$J/\psi$ and $O$ intermediate states.
Since
\begin{equation}
\left[ \begin{array}{l}
           \vert J/\psi\rangle \\
           \vert O\rangle
        \end{array}   \right]
\simeq 
\left( \begin{array}{rr}
          + c_\theta & s_\theta \\ 
           - s_\theta & c_\theta
\end{array}  \right)
\left[ \begin{array}{l}
           \vert c\bar c\rangle \\
           \vert ggg\rangle
\end{array}  \right],
\end{equation}
the factors $c_\theta s_\theta$ and $-s_\theta c_\theta$ have the same strength,
hence the total cross section under the $J/\psi$ and $O$ peaks should be the same. 
However, the $J/\psi$ is an extremely narrow state.
So long that $O$ is sufficiently broad, say a few MeV in width,
one would not expect to see an effect in $e^+e^-\to \rho\pi$ energy scan
around $m_{J/\psi}$ even if $O$ and $J/\psi$ are rather degenerate. 
For $m_O > m_{J/\psi}$, it could easily be hidden in the radiative tail.

The prognosis is therefore that eq. (1) is alive and well.
Further arguments \cite{Hou} utilize $\Gamma(J/\psi\to \rho\pi)\simeq 1.1$ MeV and
BR$(J/\psi\to \rho\pi) \simeq 1.3\%$ to constrain the  $J/\psi$-$O$
mixing angle $s_\theta$ to be $\simeq (0.025\ \rm{GeV})/((m_{J/\psi}^2 - m_O^2)^2
                                                       + m_O^2\Gamma_O^2/4)^{1/4}$.
Taking $m_O \simeq 3180$ MeV and $\Gamma_O < 50$ MeV, we find
that $f(m_{J/\psi}^2) \simeq 0.018$ GeV$^2$, which is indeed much smaller than
$f(m_{\phi}^2) \simeq 0.5$ GeV$^2$,
and $s_\theta \simeq 0.034$, which is indeed rather small ($\theta \simeq 2^\circ$).
This last point completes our argument that the vector glueball state $O$
is very clean.

\vskip 0.2cm
\noindent{\bf \boldmath{$\rhd$} Remark on OZI Violating $p\overline p\to \phi/\omega +X$}

Before closing, let us mention some possibly related phenomena in
$p\bar p$ annihilations {\it at rest}.
Define the ratio
$R_X \equiv \sigma(\bar p p \to \phi + X)/\sigma (\bar p p \to \omega + X)$.
Since $\phi$ production is OZI suppressed, one generally expects
$R_X \sim \tan^2\delta \ll 1\%$, where $\delta$ is the deviation from
ideal $\omega$-$\phi$ mixing.
This expectation is realized in many modes,
{\it except two:} $R_\gamma \sim 0.24$ and $R_\pi \sim 0.1$.
These modes are known to occur in initial $^1S_0$ and $^3S_1$ states,
respectively.
\noindent It is tempting to assume that in these two processes,
besides ``shedding" the $\gamma$ or $\pi^0$, the $p\bar p$ system
completely annihilates into three gluons and is sensitive to the
$O$ resonance.
Assuming $O$ dominance, one would find
$R_X \simeq (1/\sqrt{2})^2 = 1/2$.
Since more channels can contribute to $\omega$ production,
it leads to some reduction of $R_X$,
especially for pion case.
Unfortunately, it is not easy to quantify the arguments given here.

\vskip 0.6cm
\noindent{\Large\bf 5. 
Summary}
\vskip 0.1cm

Spectroscopy and lattice have converged on the $0^{++}$ scalar glueball $G$
having mass $m_G \simeq 1400 - 1700$ MeV,
which is considerably larger than in early 1980's.
This suggests the $1^{--}$ vector glueball $O$ to have mass 
$m_O \simeq 3$ GeV $\simeq m_{J/\psi}$.
Taking a {\it constituent} gluon picture, 
$O$ is a 3-g resonance and cannot be made from two gluons.
Its heavy mass and the 3-g content leads to suppressed mixings with
vector $q\bar q$ mesons, and a coherent picture emerges
which explains the near ideal $\omega$-$\phi$ mixing, among other things.
Proximity of $m_O \simeq  m_{J/\psi}$ (and $m_{P_3} \simeq  m_{\eta_C}$)
affect $J/\psi$ decay and explains the observed
$J/\psi ,\ \psi^\prime \to \rho\pi$ anomaly as well as similar observations
in vector-tensor modes.
These in turn fix $O$ properties:
we expect $O \to \rho\pi,\ K^*\bar K$ to be of order
1 and 0.7 MeV, with branching ratios of order 10\%,
while $O \to  p\bar p,\ K\bar K$ and $e^+e^- \sim$
10 keV, 6 keV and 6 eV are very suppressed, especially the latter.
The $O$ width is relatively narrow but cannot be extremely narrow
because of the non-observation of $O$ in a recent BES energy scan for 
$J/\psi \to \rho\pi$, and should be between 4 and 50 MeV.
Such a heavy isoscalar vector meson,
with its small width and {\it extremely suppressed $e^+e^-$ mode},
if observed, would be unmistakeable as a glueball.
This $O$ state may also be behind the extremely interesting 
``enhancement" of  $p\bar p\to \phi\gamma,\ \phi\pi$ modes against
$\omega\gamma,\ \omega\pi$ modes

\end{document}